\documentclass{elsart}

\usepackage{cite}
\usepackage{graphicx,amssymb}
\input{Qcircuit.tex}
\makeatletter
\@addtoreset{equation}{section}
\renewcommand{\theequation}{\thesection.\@arabic\c@equation}
\makeatother
\numberwithin{equation}{subsection}
\include{Qcircuit.tex}
\usepackage{latexsym}
\usepackage{amsmath}

\begin{document}

\begin{frontmatter}

\title{An Alternative Quantization Protocol for the History Dependent Parrondo Game}

\author[PsuMath]{Faisal Shah Khan\corauthref{cor}}
\corauth[cor]{Corresponding author.}
\ead{faisal@pdx.edu \\
Phone: (503) 725-3621 \\
Fax: (503) 725-3661}
\address[PsuMath]{Portland State University, Department of Mathematics and Statistics,
Portland, Oregon 97207-0751, USA}

\begin{abstract}
Earlier work on the quantization of the history dependent (HD) Parrondo game by Flitney, Ng, and Abbott led to the FNA protocol. We propose an alternative quantization protocol for this game which differs from the FNA protocol in various aspects. 
\end{abstract}

\end{frontmatter}

\section{Introduction}


The usual goal of quantizing games is to observe a different behavior in the expected payoff to the players, which should ideally be an enhancement of the values the expected payoff can take on classically. To this end, quantum games are represented by quantization protocols. The FNA (Flitney, Ng, and Abbott) quantization protocol for the history dependent (HD) Parrondo game proposed in \cite{Flitney:02} is based on maximal entanglement between the qubits. Although the authors of FNA suggest that other initial states my be used, they are neither specific nor explicit about these other initial states. Picking up on their suggestion, we point out an alternative quantization of the HD Parrondo game which assumes an initial state without entanglement. Upon measurement, the final state expresses the probability of winning by twenty one real parameters versus only four in the classical game. As in \cite{Flitney:02}, this extra dimensionality of the winning probability function allows the expected payoff to behave differently than the expected payoff in the classical game. Moreover, when the initial state is an equal superposition and the phase angles are equal, a sequence of $n$ quantum HD Parrondo games exhibits the Parrondo effect, where unlike the FNA protocol, our sequence consists of the compositions of the quantization operators.  

\section{Parrondo Games}\label{PG}

The {\it Parrondo game of type A} is a one player biased coin flipping game in which the payoff (in appropriate utility units) to the player is the expected value of the random variable representing the amount won or lost on each toss of the coin. 

For example, define a coin such that the probability of heads occurring on a toss of this coin is $p(H)=\frac{1}{2}-\epsilon$ with $\epsilon>0$. A player loses a dollar if the coin lands heads, and earns a dollar otherwise. Then clearly $P(H)<\frac{1}{2}$ and in the long run the player can expect a negative payoff. This game is biased toward losing, and can be made fair or winning by setting $\epsilon=0$ and $\epsilon<0$ respectively.

The {\it history dependent (HD) Parrondo game}, introduced in \cite{Parrondo:99} by Parrondo et al, is a one player biased coin flipping game, where the choice of the biased coin depends on the history of the game. A history dependent Parrondo game with two historical steps is reproduced here in the following table. 

\begin{center}
\begin{tabular}{c|c|c|c|c}
 Before last & Last& Coin & Prob. of win & Prob. of loss \\
$t-2$        & $t-1$& ~ & at $t$ & at $t$ \\\hline
loss & loss & $B_1'$ & $p_1$ & $1-p_1$\\ loss & win & $B_2'$ & $p_2$
& $1-p_2$ \\ win & loss & $B_3'$ & $p_3$ & $1-p_3$ \\ win & win &
$B_4'$ & $p_4$ & $1-p_4$ \\
\end{tabular}
\end{center}

A Markovian analysis in \cite{Parrondo:99} shows that the probability to win in a
generic run of a type $B$ Parrondo game is

\begin{equation}\label{p of win}
p_{\rm win}=\sum_{j=1}^{4}\pi_{st,j}p_{j}=\frac{p_1\left(p_2+1-p_4\right)}{\left(1-p_4\right)\left(2p_1+1-p_3\right)+p_1p_2}
\end{equation}

where $\pi_{st,j}$ is the probability that a certain history $j$, represented in binary format, will occur, while $p_{j}$ is the probability of win upon the flip of the third coin corresponding to history $j$. This can be rewritten as $p_{\rm win} =1/(2+c/s)$, with
$s=p_1(p_2+1-p_4)>0$ for any choice of the rules, and
$c=(1-p_4)(1-p_3)-p_1p_2.$ Therefore, game $B$ obeys the following
rule: if $c < 0$, $B$ is winning or has positive expected payoff; if $c= 0$, $B$ is
fair; and if $c> 0$, $B$ is losing or has negative expected payoff. 

The authors of \cite{Kay:03} and \cite{Parrondo:99} show the somewhat remarkable result that a random or deterministic sequence of Parrondo games of type $A$ and/or $B$ in which the individual games are biased toward losing, can be made winning. An analysis of the values of the probabilities $p$ for a game of type $A$ and probabilities $p_{i}, 1\leq i \leq 4$ for a game of type $B$ determines the {\it Parrondo region}, a region inside the unit hypercube in which the Parrondo effect manifests itself. A sequential play of Parrondo games of type $A$ and $B$ or just of game $B$ exhibits the {\it Parrondo effect} if when played individually, the two games give a negative expected payoff, but a periodic or random play of the two games leads to a positive expected payoff. 

\section{The FNA Protocol for Quantum Parrondo Games}\label{QPG}

\begin{figure}\centerline{
 \Qcircuit @C1em @R=.8em {
&\ctrlo{+1} &\qw &\ctrlo{+1} &\qw &\ctrl{+1} &\qw &\ctrl{+1} &\qw \\
  &\ctrlo{+1} &\qw &\ctrl{+1} &\qw &\ctrlo{+1} &\qw &\ctrl{+1} &\qw\\
  &\gate{g_0} &\qw &\gate{g_1} &\qw &\gate{g_2} &\qw &\gate{g_3} &\qw \\}}
\caption{\footnotesize{A quantum circuit representation of the block diagonal unitary operator used in the FNA protocol.}}
\label{3rd ord mult}
\end{figure}
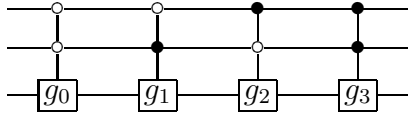

The FNA protocol for the type $A$ Parrondo game involves a qubit acted upon by an element of $SU(2)$. This is equivalent to flipping a biased quantum coin. The quantization protocol for the quantum HD game composed of two histories consists of three qubits of which the first two represent the history. Depending on the configuration of the history qubits, one out of the four possible quantum games of type $A$ is played. This protocol corresponds to an element of $SU(2^3)=SU(8)$ that has a block diagonal matrix representation with $SU(2)$ elements on each diagonal block. The block diagonal structure allows for the realization of history dependent structure of the game. A quantum circuit description of the quantization protocol for the HD game is given in figure \ref{3rd ord mult}.

For the HD game, the FNA protocol takes the initial state to be the maximally entangled state 
$$\frac{1}{\sqrt{2}}\left(\left|00 \dots 0\right\rangle+\left|11 \dots 1\right\rangle\right).
$$
Flitney et al suggest that other initial state my be used, but are not specific about what these initial states might be. We present a quantization protocol of the HD Parrondo game which uses the same unitary operator as the FNA, namely an arbitrary block diagonal $SU(8)$ element, but uses an arbitrary initial state that is a un-entangled. We describe our protocol in the following section. 

\section{Alternative Quantization Protocol for HD Parrondo Game }

Following the authors of \cite{Flitney:02}, we restrict our attention to the details of the HD game with only 2 histories. The arbitrary case for $n-1$ histories can be got easily by generalization, though the details get fairly messy. 

The classical game is embedded in the quantized game via identification of outcomes of the classical game with the basis states of the complex projective Hilbert space $\mathbb{H}^{\otimes 3}$ of qubits. We fix the basis of $\mathbb{H}^{\otimes 3}$ to be the ordered computational basis 
$$
\mathcal{B}=\left\{ \ket{000}, \ket{001}, \ket{010}, \ket{011}, \ket{100}, \ket{101}, \ket{110}, \ket{111} \right\}
$$
We take $\ket{0}$ to represent a loss and $\ket{1}$ to represent a win. The first two qubits in an element of $\mathcal{B}$ represent the history of the game, while the last qubit represents the outcome of at the present stage of the game.
Consider the subset $\mathcal{W}=\left(\ket{001}, \ket{011}, \ket{101}, \ket{111}\right)$ of $\mathcal{B}$ in which the final qubit is always in the state $\ket{1}$. The elements of $\mathcal{W}$ represent a winning outcome in the classical game.

As in the FNA, the matrix for the quantized HD Parrondo game is
\begin{equation}\label{eqn:eq15}
X=\left( {{\begin{array}{*{20}c}
 {X_1 } \hfill & 0 \hfill & 0 \hfill & 0 \hfill \\
 0 \hfill & {X_2 } \hfill & 0 \hfill & 0 \hfill \\
 0 \hfill & 0 \hfill & {X_3} \hfill & 0 \hfill \\
 0 \hfill & 0 \hfill & 0 \hfill & {X_4 } \hfill \\
\end{array} }} \right)
\end{equation}
with 
\begin{equation}\label{eqn:eq16}
X_{j}=\left( {{\begin{array}{*{20}c}
a_j \hfil & b_j \hfill \\
-\overline{b}_j \hfill & \overline{a}_j \hfill \\
\end{array}} } \right)
\end{equation}

and $a_j, b_j \in \mathbb{C}$ satisfying $\left|a_j\right|^2+\left|b_j\right|^2=1$.

However, unlike FNA, we take the initial state of three qubits to be un-entangled, say 
\begin{equation}\label{unentangled state}
\left|q_1q_2q_3\right\rangle= \left( {{\begin{array}{c}
q_{11} \\
q_{12} \\
\end{array}} } \right) \otimes \left( {{\begin{array}{c}
q_{21} \\
q_{22} \\
\end{array}} } \right) \otimes \left( {{\begin{array}{c}
q_{31} \\
q_{32} \\
\end{array}} } \right);  \hspace{.2in} \left|q_{k1}\right|^2+\left|q_{k2}\right|^2=1; \hspace{.2in} k=1,2,3.
\end{equation}

We point out here that if the initial state of the qubits was the maximally entanglement state, then a single play of the quantum HD game will {\it not} preserve the classical game due to destruction of the histories of the game. The initial state may be written as a vector in $\mathbb{H}^{\otimes 3}$.
\begin{equation}\label{vector form}
\left|q_1q_2q_3\right\rangle= \left( {{\begin{array}{c}
q_{11}q_{21}q_{31}\\
q_{11}q_{21}q_{32}\\
q_{11}q_{22}q_{31}\\
q_{11}q_{22}q_{32}\\
q_{12}q_{21}q_{31}\\
q_{12}q_{21}q_{32}\\
q_{12}q_{22}q_{31}\\ 
q_{12}q_{22}q_{32}
\end{array}} } \right); \hspace{.2in}  \sum_{r,s,t=1}^2\left|q_{1r}q_{2s}q_{3t}\right|^2=1
\end{equation}

The output from the quantum HD Parrondo game for the input state in expression (\ref{vector form}) is 

\begin{equation}\label{output}
\left|q_1q_2q_3\right\rangle= \left( {{\begin{array}{c}
q_{11}q_{21}\left(a_1q_{31}+b_1q_{32}\right)\\
q_{11}q_{21}\left(\overline{a}_1q_{32}-\overline{b}_1q_{31}\right)\\
q_{11}q_{22}\left(a_2q_{31}+b_2q_{32}\right)\\
q_{11}q_{22}\left(\overline{a}_2q_{32}-\overline{b}_2q_{31}\right)\\
q_{12}q_{21}\left(a_3q_{31}+b_3q_{32}\right)\\
q_{12}q_{21}\left(\overline{a}_3q_{32}-\overline{b}_3q_{31}\right)\\
q_{12}q_{22}\left(a_4q_{31}+b_4q_{32}\right)\\ 
q_{12}q_{22}\left(\overline{a}_4q_{32}-\overline{b}_4q_{31}\right)
\end{array}} } \right) 
\end{equation}

with the normalizing condition taking the form
\begin{eqnarray*}
&\left|q_{11}\right|^2\left(\sum_{s=1}^2\left|q_{2s}\left(a_sq_{31}+b_sq_{32}\right)\right|^2+\left|q_{2s}\left(\overline{a}_sq_{32}-\overline{b}_sq_{31}\right)\right|^2\right) \\  &+\left|q_{12}\right|^2\left(\sum_{s=1}^2\left|q_{2s}\left(a_{s+2}q_{31}+b_{s+2}q_{32}\right)\right|^2+\left|q_{2s}\left(\overline{a}_{s+2}q_{32}-\overline{b}_{s+2}q_{31}\right)\right|^2\right)=1
\end{eqnarray*} 

Note that the output or final state in (\ref{output}) exhibits entanglement between the qubits. The probability of winning, $p_{\rm{win}}^Q$ is the sum of the amplitudes of the coefficients of the elements of $\mathcal{W}$ in (\ref{output}), namely,
\begin{equation}\label{p of win}
p_{\rm{win}}^{Q}=\left|q_{11}\right|^2\left(\sum_{s=1}^2\left|q_{2s}\right|^2\left|\overline{a}_sq_{32}-\overline{b}_sq_{31}\right|^2\right)+\left|q_{12}\right|^2\left(\sum_{s=1}^2\left|q_{2s}\right|^2\left|\overline{a}_{s+2}q_{32}-\overline{b}_{s+2}q_{31}\right|^2\right)
\end{equation}  

Switching to polar form of the complex numbers and using the conditions $\left|q_{k1}\right|^2+\left|q_{k2}\right|^2=\left|a_j\right|^2+\left|b_j\right|^2=1$ for all values of $j$ and $k$, we set
\begin{equation}\label{polar qk1}
q_{k1}=e^{i\phi_{q_{k}}}\cos\left(\frac{\theta_{q_{k}}}{2}\right)
\end{equation}
\begin{equation}\label{polar qk2}
q_{k2}=e^{i\eta_{q_{k}}}\sin\left(\frac{\theta_{q_{k}}}{2}\right)
\end{equation}

\begin{equation}\label{aj}
a_j= e^{i\phi_{j}}\cos\left(\frac{\theta_{j}}{2}\right)
\end{equation}
\begin{equation}\label{bj}
b_j=e^{i\eta_{j}}\sin\left(\frac{\theta_{j}}{2}\right)
\end{equation}

with $\theta_{q_{k}},\theta_j \in \left[0,\pi\right]$, $\eta_{q_{k}},\eta_j, \phi_{q_{k}}, \phi_j \in \left[0,2\pi\right]$.  

These substitutions allow us to reduce 
\begin{equation}\label{mid term}
\left|\overline{a}_jq_{32}-\overline{b}_jq_{31}\right|^2 =  \cos^2\left(\frac{\theta_j}{2}\right)\sin^2\theta_{q_{3}}+\sin^2\left(\frac{\theta_j}{2}\right)\cos^2\theta_{q_{3}}-2\Re\left\{a_j\overline{b}_jq_{31}\overline{q}_{32}\right\}
\end{equation}

with
$$
\Re\left\{a_j\overline{b}_jq_{31}\overline{q}_{32}\right\}=\cos\left(\theta_j-\eta_j+\phi_{q_{3}}-\eta_{q_{3}}\right)\cos\left(\frac{\theta_j}{2}\right)\sin\left(\frac{\theta_j}{2}\right)\cos\theta_{q_{3}}\sin\theta_{q_{3}}
$$
Substituting equations (\ref{polar qk1}) - (\ref{mid term}) into equation (\ref{p of win}) allows us to express $p_{\rm{win}}^Q$ in terms of {\it twenty one} real variables, each taking on values from the closed unit interval, versus only four such variable that express $p_{\rm{win}}$. 

The expected payoff to the player in a single play of the classical HD Parrondo game is $2p_{\rm{win}}-1$ while in the quantum game it is $2p_{\rm{win}}^Q-1$. Quantization will enhance the payoff if $p_{\rm{win}}^Q > p_{\rm{win}}$. However, it is possible that for some range of values of some of the domain variables $p_{\rm{win}}^Q \leq p_{\rm{win}}$. In other words, the choice of initial state and the choice of the values of the phase variables in the unitary operator for the quantization lends the expected payoff a different behavior than the expected payoff in the classical game. 

\subsection{Equal Superposition as the Initial State}

Consider the following case in which the initial state is the equal superposition
\begin{equation}\label{equal super}
\left( {{\begin{array}{c}
\frac{1}{\sqrt{2}} \\
\frac{1}{\sqrt{2}} \\
\end{array}} } \right) \otimes 
\left( {{\begin{array}{c}
\frac{1}{\sqrt{2}} \\
\frac{1}{\sqrt{2}} \\
\end{array}} } \right) \otimes 
\left( {{\begin{array}{c}
\frac{1}{\sqrt{2}} \\
\frac{1}{\sqrt{2}}\\
\end{array}} } \right)=\frac{1}{\sqrt{8}}\left(1,1,1,1,1,1,1\right)^{T}.
\end{equation}
Then $p_{\rm{win}}^Q$ is simply

\begin{equation}\label{eqn19}
p_{\rm{win}}^{Q}=\frac{1}{8}\left(4-\sum_{j=1}^{4}\sin\theta_{j}\cos(\eta_j-\phi_j)\right)=\frac{1}{2}-\frac{1}{8}\sum_{j=1}^{4}\sin\theta_{j}\cos(\eta_j-\phi_j)
\end{equation}

For values of the phase angles $\eta_j, \phi_j$ such that $\sum_{j=1}^{4}\sin\theta_{j}\cos(\eta_j-\phi_j) >0$, the quantum game is losing. For values that satisfy $\sum_{j=1}^{4}\sin\theta_{j}\cos(\eta_j-\phi_j)<0$, the quantum game is winning, where as the game is fair when $\sum_{j=1}^{4}\sin\theta_{j}\cos(\eta_j-\phi_j) =0$.

The simplest case to analyze is when $\eta_j=\phi_j$ since in this case 
$$
p_{\rm{win}}^{Q}=\frac{1}{2}-\frac{1}{8}\sum_{j=1}^{4}\sin\theta_{j}
$$
and is equal to $\frac{1}{2}$ for $\theta_j=0$ or $\theta_j=\pi$, meaning that the quantum game is fair. However, $p_{\rm{win}}^{Q}$ is equal to $\frac{1}{2}-\epsilon$, $\epsilon > 0$, for all other values of the $\theta_j$, meaning that the game is losing. Next we see how a sequence of such games exhibits the Parrondo effect.

\subsection{Parrondo Effect}
Our definition of a sequence of quantum HD games differs from the FNA definition as follows.

A sequence of $n$ three qubit quantum HD games in the FNA protocol is played with $3n$ qubits, with each game in the sequence shifting down one qubit. For us, a sequence of $n$ three qubit quantum HD games is the composition of two quantum multiplexers, and therefore the number of qubits is always three. 

In the present example with the initial state being the equal superposition, $p_{\rm{win}}^{Q}$ for the sequence of $n$ games is just

$$
p_{\rm{win}}^{Q}=\frac{1}{2}-\frac{1}{8}\sum_{j=1}^{4}\sin\left(\sum_{k=1}^{n}\theta_{j}^{k}\right)
$$

The term $\sum_{k=1}^{n}\theta_{j}^{k}$ enlarges the range of the sine functions from $\left[0,\pi\right]$ to $\left[0,n\pi\right]$, allowing the expression $\sum_{j=1}^{4}\sin\left(\sum_{k=1}^{n}\theta_{j}^{k}\right)$ to take on negative values and for $p_{\rm{win}}^{Q}=\frac{1}{2}+\epsilon$ with $\epsilon >0$.

\section{Conclusion}

We proposed an alternative quantization for the history dependent Parrondo game in which the initial state is un-entangled, yet, due to the presence of extra parameters, the expected payoff can be made to behave differently than the expected payoff in the classical game. We give a simple  example of this behavior with three qubits, where the initial state of qubits has no entanglement, and the phase angles equal. Because the phase angles contribute nothing to the probability of win, this example is analogous to the classical HD Parrondo game. However, in this case, the quantum game is a losing game! For this simple example, we also established that the the Parrondo effect is present in the play of a sequence of $n$ games. 

The extra dimensionality of the expected payoff from the quantum game can the topic of future research. Clearly, twenty one variables in the winning probability function in the quantum game enlarge the range in which the game can be won or lost, and in which the Parrondo effect can manifest. 

We point out an interesting fact that will be a topic of our future research. The unitary operator used in the FNA and in this document to quantize the HD Parrondo game is known as a quantum multiplexer in the language of quantum logic synthesis. When an arbitrary quantum circuit is synthesized using the cosine-sine synthesis (CSS) method \cite{FSK:06, shende:05}, the resulting circuit consists entirely of a sequence of quantum multiplexers of various orders. The various orders of the quantum multiplexer may be used to quantize the HD Parrondo game, in which case the CSS may be interpreted as a sequence of quantum HD Parrondo games and it is worth while to explore whether the Parrondo effect manifest in such a sequence.

\section{Acknowledgment}
The author is grateful to Steve Bleiler, Bryant York, and Aden Ahmed for discussions and suggestions. The quantum circuit diagrams were all drawn
in LATEX using Q-circuit available at http://info.phys.unm.edu/Qcircuit/.

\end{document}